\documentclass[%
11pt,
reprint,
onecolumn,
tightenlines,
superscriptaddress,
preprintnumbers,
nofootinbib,
amsmath,amssymb,amsthm,
physrev,
eqsecnum,tikz,
]{revtex4-2}

\newcommand{\RN}[1]{\textup{\uppercase\expandafter{\romannumeral#1}}}

\newcommand{\I}{\RN{1}}
\newcommand{\II}{\RN{2}}

\usepackage{isomath}
\usepackage{amsmath,amsthm}
\usepackage{amsbsy}
\usepackage{amssymb}
\usepackage{amscd}
\usepackage{amsfonts}
\usepackage{stmaryrd}
\usepackage{siunitx}
\usepackage{euscript}
\usepackage[utf8]{inputenc}
\usepackage[T1]{fontenc}
\usepackage{newtxtext} 
\everymath{\displaystyle}
\usepackage{exscale}
\usepackage{microtype}
\usepackage{hyperref}
\usepackage{booktabs}
\usepackage{algorithm}
\usepackage{algpseudocode}

\usepackage{graphicx}
\usepackage{boxedminipage}
\usepackage{calc}
\usepackage[usenames,dvipsnames]{xcolor}
\graphicspath{ {media/} }
\usepackage[caption=false,justification=centerlast]{subfig}

\usepackage{setspace}
\usepackage{enumitem}
\setitemize{noitemsep,topsep=0pt,parsep=0pt,partopsep=0pt}
\setenumerate{noitemsep,topsep=0pt,parsep=0pt,partopsep=0pt}
\setdescription{noitemsep,topsep=0pt,parsep=0pt,partopsep=0pt}

\usepackage[colorinlistoftodos, color=green!40,prependcaption]{todonotes}
\setuptodonotes{inline}

\usepackage{soul} 
\usepackage[normalem]{ulem}

\usepackage{orcidlink}
\usepackage{siunitx}
\usepackage[small]{titlesec}

\titlespacing*{\section}{0pt}{12pt plus 4pt minus 2pt}{2pt plus 2pt minus 2pt}
\titlespacing*{\subsection}{0pt}{12pt plus 4pt minus 2pt}{2pt plus 2pt minus 2pt}
\titlespacing*{\subsubsection}{0pt}{12pt plus 4pt minus 2pt}{2pt plus 2pt minus 2pt}
\titlespacing*{\paragraph}{0pt}{12pt plus 4pt minus 2pt}{2pt plus 2pt minus 2pt}

\makeatletter

    \renewcommand*{\p@subsection}{}
    
    \renewcommand*{\p@subsubsection}{}
\makeatother

\usepackage{isomath}
\usepackage{amsmath}
\usepackage{amssymb}
\usepackage{amscd}
\usepackage{amsfonts}
\usepackage{pgf}
\usepackage{import}
\usepackage[export]{adjustbox}
\usepackage[percent]{overpic}
\usepackage{circledsteps}

\pgfkeys{/csteps/inner ysep=7.5pt}
\pgfkeys{/csteps/inner xsep=7.5pt}

\newcommand{\eps}{{\varepsilon}}

\newcommand{\half}{\frac{1}{2}}
\newcommand{\hE}{\hat{E}}

\newcommand{\bfSigma}{\mathbold {\Sigma}}

\newcommand{\ga}{\gamma}

\DeclareMathOperator{\trace}{tr}

\newcommand{\parderiv}[2]{\frac{\partial #1}{\partial #2}}

\newcommand{\dm}{\ \mathrm{d}}

\newcommand{\bfe}{{\mathbold e}}

\newcommand{\bfn}{{\mathbold n}}

\newcommand{\bft}{{\mathbold t}}

\newcommand{\bfx}{{\mathbold x}}

\newcommand{\bfE}{{\mathbold E}}
\newcommand{\bfF}{{\mathbold F}}

\newcommand{\bfP}{{\mathbold P}}

\newcommand{\bfT}{{\mathbold T}}

\begin{document}


\preprint{Physical Review Applied, 23:014007, 2025 (10.1103/PhysRevApplied.23.014007)}

\title{Exploiting Instabilities to Enable Large Shape Transformations in Dielectric Elastomers}

\author{Daniel Katusele}
    \email{dkatusel@andrew.cmu.edu}
    \affiliation{Department of Civil and Environmental Engineering, Carnegie Mellon University}

\author{Carmel Majidi}
     \affiliation{Department of Mechanical Engineering, Carnegie Mellon University}

\author{Pradeep Sharma}
     \affiliation{Department of Mechanical Engineering, University of Houston}

\author{Kaushik Dayal \orcidlink{0000-0002-0516-3066}}
    \affiliation{Department of Civil and Environmental Engineering, Carnegie Mellon University}
    \affiliation{Center for Nonlinear Analysis, Department of Mathematical Sciences, Carnegie Mellon University}
    \affiliation{Department of Mechanical Engineering, Carnegie Mellon University}

\date{\today}


\begin{abstract}
    Dielectric elastomers have significant potential for new technologies ranging from soft robots to biomedical devices, driven by their ability to display complex shape changes in response to electrical stimulus.
    However, an important shortcoming of current realizations is that large voltages are required for useful actuation strains.
    This work proposes, and demonstrates through theory and numerical simulations, a strategy to achieve large and controlled actuation by exploiting the electromechanical analog of the Treloar-Kearsley (TK) instability.
    
    The key idea is to use the fact that the TK instability is a symmetry-breaking bifurcation, which implies the existence of a symmetry-driven constant-energy region in the energy landscape.
    This provides for nonlinear soft modes with large deformations that can be accessed with very small external stimulus, which is achieved here by applying a small in-plane electric field.
    First, the bifurcation and post-bifurcation behavior of the electromechanical TK instability are established theoretically in the idealized setting of uniform deformation and electric field.
    Next, building on this, a finite element analysis of a realistic geometry with patterned top and bottom electrodes is applied to demonstrate large and soft shape changes driven by small voltage differences across the electrodes.

\end{abstract}

\maketitle


\section{Introduction}


Soft electromechanical elastomers, also known as Dielectric Elastomers (DE), wherein mechanical deformations can be driven by electrical stimulus for actuation, present tremendous potential as transducers for soft and biologically-inspired robots, biomedical devices, energy harvesting, among other applications \cite{carpi2011bioinspired, rus2015design, kofod2007energy, pelrine2000highs,pelrine2000highf, bauer201425th, yang2017avoiding, koh2009maximal, bar2004electroactive, bartlett2016stretchable,grasinger2021flexoelectricity,grasinger2021architected}. 
The actuation mechanism in DE is typically achieved through a capacitor-like design where a dielectric elastomer film is sandwiched between two compliant electrode: upon application of a voltage difference across the electrodes, the electrostatic (Coulombic) force between the electrodes due to the electrical charges compresses the DE in the thickness direction causing --- through the Poisson effect --- the DE to expand in the lateral direction \cite{pelrine2000highstr, wissler2007mechanical, kollosche2012complex}.

However, because the Poisson effect is typically fairly small, DE typically require high voltages to induce a usable deformation.
Consequently, the high fields drive electromechanical failures such as pull-in instability, electrical breakdown, and buckling instability \cite{zhao2014harnessing, bense2017buckling, lu2020mechanics} which limit the performance. 
To increase the deformation and delay/suppress instabilities, mechanical prestretch and prestress are often introduced before applying voltage \cite{li2011effect, kofod2008static, yang2017avoiding, huang2012giant}. 
Other methods to improve the performance of DE include the introduction of mechanical constraints \cite{zhao2014harnessing, zhang2011mechanical}, using dielectric films without electrodes \cite{keplinger2010rontgen}, or harnessing instabilities for improved functionality \cite{huang2012giant}.

Dielectric elastomers have also been used in multilayer configuration to increase stability, hence allowing to bypass the need to apply a prestretch to the material \cite{duduta2016multilayer}.
This strategy also eliminates some electromechanical instabilities such as wrinkling that would otherwise limit the amount of deformation achievable.
Multilayer DE actuators have been applied to flapping wing robots \cite{chen2019controlled}, haptics \cite{zhao2020wearable}, and inspection robots \cite{tang2022pipeline}.

In this work, we exploit the Treloar-Kearsley (TK) instability to theoretically and computationally show the possibility of large shape transformations in DE with small applied electric field.
The TK instability is a symmetry-breaking pitchfork bifurcation that occurs in polymeric materials in the mechanical setting: under symmetric loading, a stable asymmetric stretching can be observed in addition to the unstable symmetric stretching \cite{treloar1948stresses,kearsley1986asymmetric,steigmann2007simple,batra2005treloar}. 
This was recently studied in the electromechanical context in DE to elucidate the interplay between the TK instability and pull-in instability under applied voltage \cite{chen2021interplay}.

\paragraph*{Contributions of this paper.}

\begin{figure*}[htb!]
    \centering
    \includegraphics[width=0.45\textwidth]{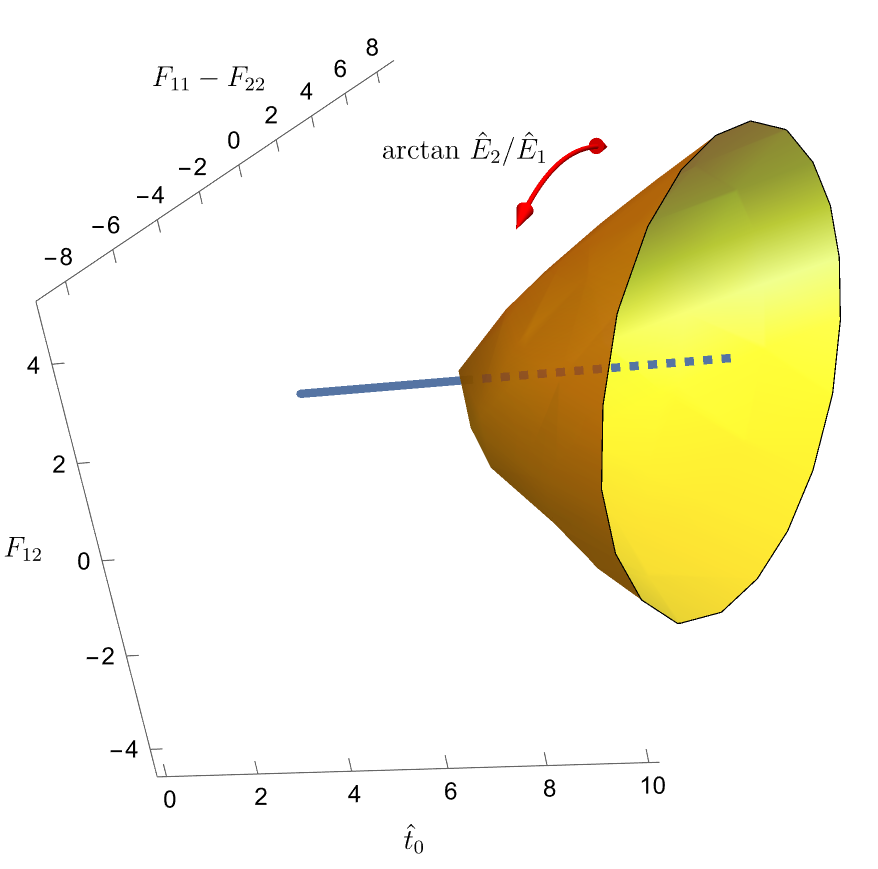}
    \caption{
        Bifurcation schematic illustrating the TK pitchfork instability. 
        $F_{12}$ and $F_{11}-F_{22}$ quantify the shape of the cross-section, and $\hat{t}_0$ quantifies the mechanical load. 
        For low loads, the stable configuration is circular, shown by the solid line. 
        For large loads, the circular configuration is unstable, shown by the dashed line, and the surface of revolution is the stable configuration that corresponds to a large set of constant energy configurations.
        It is then possible, for a fixed load, to move along circular paths on the surface of revolution for zero energetic cost --- i.e., a soft mode --- simply by using the orientation $\arctan \hat{E}_2/\hat{E}_1$ of a small in-plane electric field to break the symmetry.
    }
\label{fig:idea}
\end{figure*}

This paper proposes a strategy to exploit the symmetry-breaking feature of the TK instability to achieve large mechanical deformation with small electrical stimulus (Fig. \ref{fig:idea}).
Specifically, considering a specimen that is initially a flat circular disk, we can mechanically induce the TK instability to cause a shape change to a flat disk with elliptical cross-section.
There is clearly a broken symmetry: the principal axes of the elliptical cross-section are free to be oriented in any direction.
All directions have the same energy, and this provides a flat region in the energy landscape; specifically, once the TK instability has occurred and the cross-section is elliptical, we can continuously change the orientation of the principal axes without any energetic cost.
This flat region provides for the existence of a nonlinear soft mode, i.e., we can achieve large mechanical deformations that require no energy.
Given this soft mode in the energy, we next apply a small in-plane electric field that breaks the in-plane symmetry and nudges the system to orient the principal axes along the field.
Changing the in-plane orientation of the field causes the principal axes to follow along, thereby providing large deformations that are readily controlled by a small field.

In Section \ref{forml}, we summarize the large deformation nonlinear electromechanical model starting from a free energy formulation.
By minimizing the free energy, we obtain the partial differential equations (PDE) and boundary conditions that define the electromechanical equilibrium state.

Next, in Section \ref{bifurc}, we perform a theoretical bifurcation analysis in an idealized setting.
Assuming a homogeneous deformation and a uniform electric field applied through affine voltage boundary conditions, we simplify the full PDE model to an algebraic system.
This enables us to examine the bifurcation and post-bifurcation behavior, specifically understanding the effect of applied field on the onset of the TK instability and the post-bifurcation symmetry breaking.

In Section \ref{fem}, we solve the equilibrium PDE numerically using the finite element method (FEM) in a realistic geometry that is potentially amenable to fabrication and experimental characterization in the future.
Specifically, we consider a circular disk with patterned top and bottom electrodes that can be individually addressed in terms of applied voltage.
By changing the voltages at the various electrodes, it is possible to approximately mimic an in-plane applied electric field that has various orientations.
We demonstrate that changing the electrode voltages appropriately leads to the large and soft deformations predicted by the idealized theoretical analysis.

Finally, in Section \ref{sec:multi-layer}, we apply the numerical approach to examine a realistic multilayer DE actuator configuration.

\section{Formulation} \label{forml}

This section describes the electromechanical free energy formulation, focusing on the presentation of the model which will be analyzed in subsequent sections. 
We use boldface to represent vectors and tensors, and subscripts of $0$ to denote quantities in the reference configuration.
We have used $\nabla$ and $\nabla_0$ to denote derivatives with respect to $\bfx$ and $\bfx_0$ respectively.

\subsection{Variational Principle and Field Equations}

Consider a DE specimen occupying the domain $\Omega_0$ and boundary $\partial \Omega_0$ in the reference configuration, and $\Omega$ with boundary $\partial \Omega$ in the deformed configuration. 
The material points in the reference and deformed configurations are denoted by $\bfx_0$ and $\bfx$ respectively, and the deformation gradient by $\bfF(\bfx_0) = \nabla_0 \bfx$ consistent with the deformation map $\bfx = \bfx(\bfx_0)$; we further define $J = \det(\bfF) > 0$ as the Jacobian. 
The electric potential $\phi(\bfx)$ and electric field $\bfE(\bfx)$ are related by $\bfE = - \nabla \phi$.
The polarization field in the material is denoted as $\bfP(\bfx)$. 

The total free energy of the system is formulated \cite{liu2013energy} as 
\begin{equation} 
    \psi[\bfx, \bfP] 
    = 
    \int_{\Omega_0} W(\bfF, \bfP) + \frac{\eps_0}{2} \int_{\Omega} |\bfE|^2 - \int_{\partial \Omega_{t_0}} \bft_0 \cdot \bfx + \int_{\partial \Omega} \phi (\eps_0 \bfE + \bfP) \cdot \bfn \label{Elr_Engy}
\end{equation}
where $W$ is the free energy density per unit referential volume; the second term is the electrostatic field energy, noting that $\eps_0$ is the permittivity of free space; and the last two terms are the contributions from the  mechanical and electrical boundary conditions respectively. 
$\partial \Omega_{t_0}$ is the part of the boundary $\partial \Omega_0$ where the traction is specified with $\bfn$ the outward normal. 
The electric field in \eqref{Elr_Engy} is computed by solving the electrostatic equation :
\begin{equation}  
    \nabla \cdot (\eps_0 \bfE + \bfP) = -\eps_0 \nabla^2 \phi + \nabla \cdot \bfP = 0 \quad \text{in } \Omega \label{Maxwell}
\end{equation}
subject to the boundary conditions that $\phi$ is specified at the electrodes and $\left(\eps_0 \bfE + \bfP\right) \cdot \bfn = 0$ on the portion of the boundary where there are no free charges.
We note that this is an approximation that neglects the external electric fields outside the specimen \cite{yang2011completely,jha2023discrete}.

The pullbacks to the reference configuration of $\bfE(\bfx), \bfP(\bfx)$ are defined following \cite{marshall2014atomistic,jha2023atomic} to be:
\begin{equation} 
    \bfE_0 = \bfF^T \bfE, \quad \bfP_0 = J \bfP\nonumber 
\end{equation}
and the pullback for the electric potential to be $\phi_0(\bfx_0) = \phi(\bfx(\bfx_0))$. 

The dielectric elastomer is assumed to be incompressible, which requires that $J = 1$, and is imposed by introducing a Lagrange multiplier $p(\bfx_0)$.
The Lagrangian for an incompressible material, written in terms of the pullbacks, has the expression:
\begin{equation}
    \Psi[\bfx, \bfP_0] = \int_{\Omega_0} W(\bfF, \bfP_0) + \frac{\eps_0}{2} \int_{\Omega_0} J |\bfF^{-T} \bfE_0|^2 - \int_{\partial \Omega_{t_0}} \bft_0 \cdot \bfx  + \int_{\partial \Omega_0} \phi_0 J \bfF^{-1}( \eps_0 \bfF^{-T} \bfE_0 + \bfP_0)  \cdot \bfn_0 - \int_{\Omega_0} p (J-1) \label{Lgr_Engy}
\end{equation}  
where $\bfn_0$ is the outward normal to $\partial \Omega_0$. 

Setting the functional derivative of $\Psi$ with respect to $\bfx(\bfx_0)$ to $0$, with the constraint \eqref{Maxwell}, we obtain the following equations that represent mechanical equilibrium in the bulk and the boundary conditions:
\begin{subequations}
\begin{align} 
    \nabla_0 \cdot \left( \frac{\partial W}{\partial \bfF} + \bfSigma_0 - p J \bfF^{-T}\right) = \mathbf{0}  & \quad \text{on } \Omega_0 \label{equil1} \\
    \left( \frac{\partial W}{\partial \bfF} + \bfSigma_0 - p J \bfF^{-T}\right) \bfn_0 = \bft_0 & \quad \text{on } \partial \Omega_{t_0} \label{equil2} \\
    \left( \frac{\partial W}{\partial \bfF} + \bfSigma_0 - p J \bfF^{-T}\right) \bfn_0 = \mathbf{0} & \quad \text{on } \partial \Omega_0 \setminus \partial \Omega_{t_0} \label{equil3}
\end{align}
\label{eqn:equil}
\end{subequations}
We have defined $\bfSigma_0 := \bfE_0 \otimes J \bfF^{-1}( \eps_0 \bfF^{-T} \bfE_0 + \bfP_0) - \frac{\eps_0 J}{2} |\bfE_0|^2 \bfF^{-T}$ as the Piola-Maxwell stress tensor, and $\bfT := \frac{\partial W}{\partial \bfF} + \bfSigma_0 - p J \bfF^{-T}$ as the total Piola-Kirchhoff stress tensor.
The PDE and BCs in \eqref{eqn:equil} define the boundary value problem (BVP) that must be solved for the equilibrium configuration.

Similarly, setting the functional derivative of $\Psi$ with respect to $\bfP_0(\bfx_0)$ to $0$ gives the usual local relation between the electric field and polarization density at a point $-\parderiv{W}{\bfP_0} = \bfE$ \cite{darbaniyan2019designing}.

\subsection{Material Model}

We assume that the energy density $W(\bfF, \bfP_0)$ is additively composed of a mechanical strain energy density $W^m(\bfF)$ and an electromechanical energy density $W^{el}(\bfF, \bfP_0)$. 

For the mechanical term, we use an incompressible, isotropic, hyperelastic Mooney-Rivlin model \cite{rivlin1948large}, that can be connected to statistical mechanics and network elasticity \cite{khandagale2023statistical,grasinger2023polymer}, with the form:
\begin{equation} 
    W^m(\bfF) = \frac{\mu}{2}\left[\left(\I_1 - 3 \right) + \ga \left( \I_2 - 3 \right) \right]  \label{strain-engy}
\end{equation}
where $\mu$ and $\ga$ are material parameters, and $\I_1  = \trace (\bfF^T \bfF)$ and $\I_2 = \half \left(\trace (\bfF^T \bfF)^2 - \trace ((\bfF^T \bfF)^2)\right)$ are the invariants of $\bfF$. 
All quantities will be non-dimensionalized with respect to $\mu$ and we set $\ga=0.3$ for all numerical calculations.

For the electromechanical term, we use a linear isotropic dielectric with $\bfP = \eps_0 \chi \bfE$, where $\chi$ is the scalar dielectric susceptibility. 
In \cite{grasinger2021nonlinear}, it was shown that the dielectric susceptibility derived from statistical mechanics is an anisotropic function of the deformation; however, for simplicity, we assume that $\chi$ is isotropic and independent of deformation. 
Defining $\eps := \eps_0(1+ \chi)$ to be the permittivity, we write the electromechanical energy as \cite{liu2014energy}:
\begin{align} \label{pol}
    W^{el} = \frac{1}{2J} \bfP_0 \cdot (\eps - \eps_0)^{-1} \bfP_0.
\end{align}
We note that though $\eps$ is independent of deformation, $W^{el}$ involves the deformation through the presence of $J=\det \bfF$ and because $\bfP_0$ depends on $\bfF$ through the pullback relation.

\section{Bifurcation and Post-Bifurcation Analysis in an Idealized Homogeneous System} \label{bifurc}

In this section, we examine the TK bifurcation in an idealized setting --- with deformation and electric field being uniform through the specimen --- to simplify the partial differential equation \eqref{eqn:equil} to an algebraic system of equations.
The algebraic system can be solved to obtain closed-form expressions that elucidate the role of mechanical load and electric field orientation in the electromechanical TK bifurcation.


\subsection{Simplification of the Differential Equation to an Algebraic System}

The necessary conditions for the onset of the symmetry-breaking instability is determined through a linear bifurcation analysis on a DE specimen subject to both mechanical loads and electrical stimuli. 
We consider a disk-shaped specimen, and the deformation and electric field are both assumed to be homogeneous under the applied loads.
This enables us to simplify our analysis for TK and pull-in instabilities, but restricts it to situations without buckling instabilities.

The Cartesian coordinates of material points in the reference configuration are of the form $\bfx_0 = (x_1, x_2, x_3) = (R \cos \theta, R \sin \theta, x_3)$, where $R$ is the radius of the disk and $\theta\in [0,2\pi)$.
The corresponding spatial position of these material points after deformation is of the form $\bfx = (F_{11} x_1 + F_{12} x_2, F_{21} x_1 + F_{22} x_2, F_{33} x_3)$, with 
\begin{equation} 
    \bfF 
    = 
    \begin{pmatrix}
        F_{11} & F_{12} & 0\\
        F_{21} & F_{22} & 0 \\
        0 & 0 & F_{33}
    \end{pmatrix} 
\end{equation}
The components $F_{13}, F_{23}, F_{31}$ and $F_{32}$ are negligible because the specimen has a thickness that is small compared to its radius.

The mechanical load is applied uniformly on the entire lateral boundary by specifying the traction $\bfT \bfe_r = \bft_0$, with $\bft_0 = t_0 \bfe_r$ and $\bfe_r = (\cos{\theta}, \sin \theta, 0)$.
The top and bottom faces are traction free, i.e $\bfT \bfe_3 = \bf0$. 
The term $\bft_0 \cdot \bfx$ in \eqref{Lgr_Engy} evaluates to $Rt_0((F_{11}x_1 + F_{12}x_2)\cos \theta + (F_{21}x_1 + F_{22}x_2)\sin \theta))$. 
The energy contribution due to the applied traction can now be written as
\begin{align}
    \int_{\partial \Omega_{t_0}} \bft_0 \cdot \bfx = RHt_0 \int_0^{2 \pi} (F_{11}\cos^2 \theta + F_{22} \sin^2 \theta + (F_{12} + F_{21}) \cos \theta \sin \theta)R \dm \theta = \pi R^2Ht_0 (F_{11} + F_{22})
\end{align}

The voltage boundary condition has the affine form $\phi = -\bfE^{ext} \cdot \bfx$ and is applied on the entire boundary $\partial\Omega$, where $\bfE^{ext} = (E_1, E_2, E_3)$ is a constant vector with the physical interpretation of a uniform applied electric field \cite{grasinger2020statistical,grasinger2022statistical,khandagale2024statistical}.
The internal electric field that is generated by this boundary voltage is computed from \eqref{Maxwell} and evaluates to $\nabla \phi = -\bfE^{ext}$. 

The energy due to the electric field in \eqref{Lgr_Engy}, by using \eqref{pol} and under the assumption of a homogeneous deformation, simplifies to:
\begin{equation}
    \frac{\eps}{2} \int_{\Omega_0} J |\bfF^{-T} \bfE^{ext}|^2 + \int_{\partial \Omega_0} \phi_0 \bfE^{ext}  \cdot \bfn_0 = -\pi R^2 H\frac{\eps}{2} |\bfF^{-T} \bfE^{ext}|^2 
\end{equation}
where we have used the prescribed affine voltage boundary conditions and the divergence theorem.

Using all the simplified expressions from above, the mean free energy, in terms of components of $\bfF$ and $\bfE^{ext}$ can be written as:
\begin{equation}
\begin{split}
    \frac{1}{\pi R^2H}\psi[F_{11}, F_{12}, F_{21}, F_{22}, F_{33}] 
    = 
    & 
    \frac{\mu}{2}\left((\I - 3) + \ga( \II - 3)\right) - t_0(F_{11} + F_{22}) 
    - \frac{\eps}{2} (F_{22}E_1^{ext} - F_{21}E_2^{ext})^2F_{33}^2 
    \\
    & 
    - \frac{\eps}{2} (F_{11}E_2^{ext} - F_{12}E_1^{ext})^2F_{33}^2 - \frac{\eps}{2} \frac{(E_3^{ext})^2}{F_{33}^2} - p \left(F_{11}F_{22}F_{33}-F_{12}F_{21}F_{33} -1\right)
\end{split}
 \label{eq:comp_engy}
\end{equation}
This generalizes the expression from \cite{chen2021interplay} which was restricted to out-of-plane electric fields.

\subsection{Analysis}

Unlike the classical purely mechanical analysis \cite{kearsley1986asymmetric}, we cannot assume here that $\bfF$ is symmetric and diagonalizable in terms of principal stretches; in fact, we find below that $\bfF$ is generally not symmetric when there is an in-plane electrical field.
That is, the initially-circular specimen deforms to a more general shape than an ellipsoid, though it is very close to an ellipsoid for the realistic values of the electric field that we use here.
We therefore cannot use the strategy of working in the principal basis of $\bfF$, and instead choose a Cartesian coordinate system that, without loss of generality, is oriented such that $E_2 = 0$.

The linear bifurcation analysis is carried out by solving the equilibrium equations \eqref{eqn:equil} under the assumption of homogeneous deformations --- which leads to the field equation \eqref{equil1} being trivially satisfied --- with the total energy is given by \eqref{eq:comp_engy}. 
The boundary conditions \eqref{equil3} with zero traction on the top and bottom faces lead to the normal stress component in the thickness direction being zero, i.e., $T_{33}=0$, and enables us to eliminate the pressure $p$.
Further, the in-plane stress components $T_{11}$, $T_{12}$, $T_{21}$ and $T_{22}$ associated with the boundary condition \eqref{equil2}, in which the normal corresponds to the radial directions, leads to:
\begin{align} 
    \label{str1}
    T_{11} = t_0, ~ T_{12} = 0, ~T_{21} = 0, ~\text{and}~T_{22} = t_0.
\end{align}

To simplify, we compute $T_{11} - T_{22}=0$ and $T_{12} - T_{21}=0$ to get respectively:
\begin{subequations}
\begin{equation}
       -\hE_1^2 F_{22} F^{-1}_{33} + (F_{11}-F_{22}) \left[ \left( 2 \ga - \hE_3^2\right) F^{-4}_{33} -\left(2 \ga - \hE_1^2 \right) \left(F_{12}^2 + F_{22}^2 \right) -  
       2 + 2 F^{-3}_{33}  + 2\ga  \left(F_{11}^2 +F_{22}^2 + F^{-1}_{33} \right)  \right]=0
\end{equation}
and
\begin{equation}
    \hE_1^2 F_{12} F^{-1}_{33}+ \left( F_{12}-F_{21} \right) 
    \begin{bmatrix}
        \left( 2 \ga -\hE_1^2 \right) \left( F_{12}^2 + F_{22}^2 \right) - \left( 2 \ga -\hE_3^2 \right) F^{-4}_{33} + 2 \ga  \left( F_{11}^2 + F_{21}^2 -F^{-1}_{33}\right) \\ -2 F^{-1}_{33} \left( F^2_{11} F^2_{22} -2F_{11}F_{12} F_{21}F_{22} + F^2_{12}F^2_{21} \right)
    \end{bmatrix}
    =0
\end{equation}
\end{subequations}
where we have introduced the non-dimensional electric field $\hE_i := \frac{E^{ext}_i}{\sqrt{\mu / \eps}}, i = 1, 3$.

In this absence of electric fields, i.e., $\hE_1 = \hE_3 = 0$, this agrees with Kearsley's theoretical analysis in predicting asymmetric deformation beyond a critical load \cite{kearsley1986asymmetric}. 
Further, $\hE_1$ introduces an asymmetry in the plane of the dielectric such that $\bfF$ is not symmetric, i.e., in general, $E_1 \neq 0 \implies F_{12} \neq F_{21}$.

\subsection{Results and Discussion} \label{Analysis_Results}

\begin{figure}[htb!]
    \centering
    \subfloat[]{\includegraphics[width=0.49\textwidth]{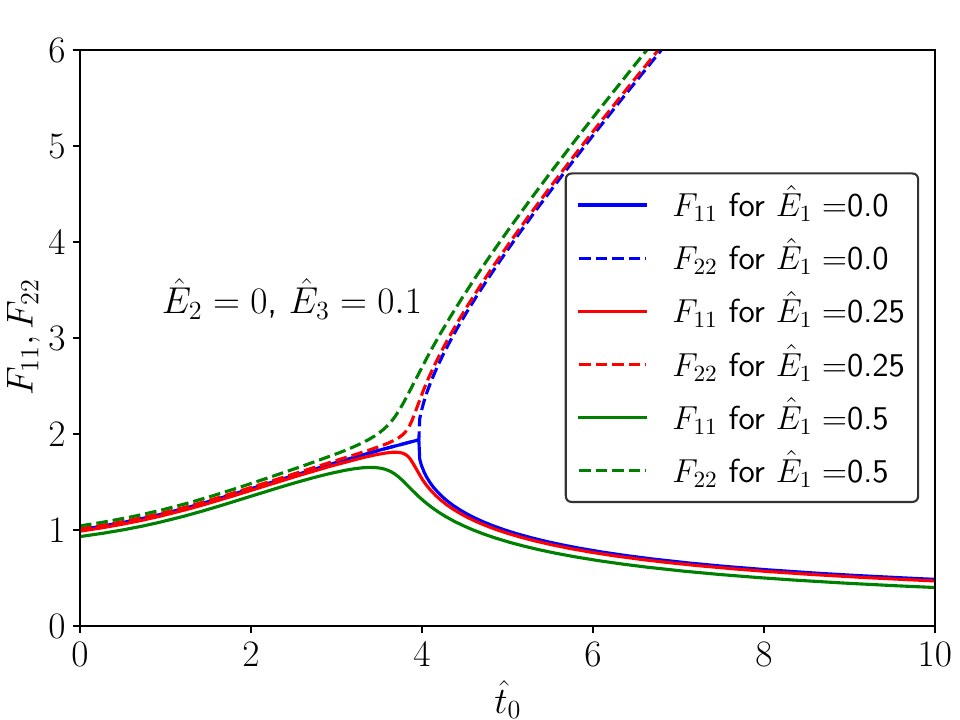}\label{fig:bif1}}
    \hfill
    \subfloat[]{\includegraphics[width=0.49\textwidth]{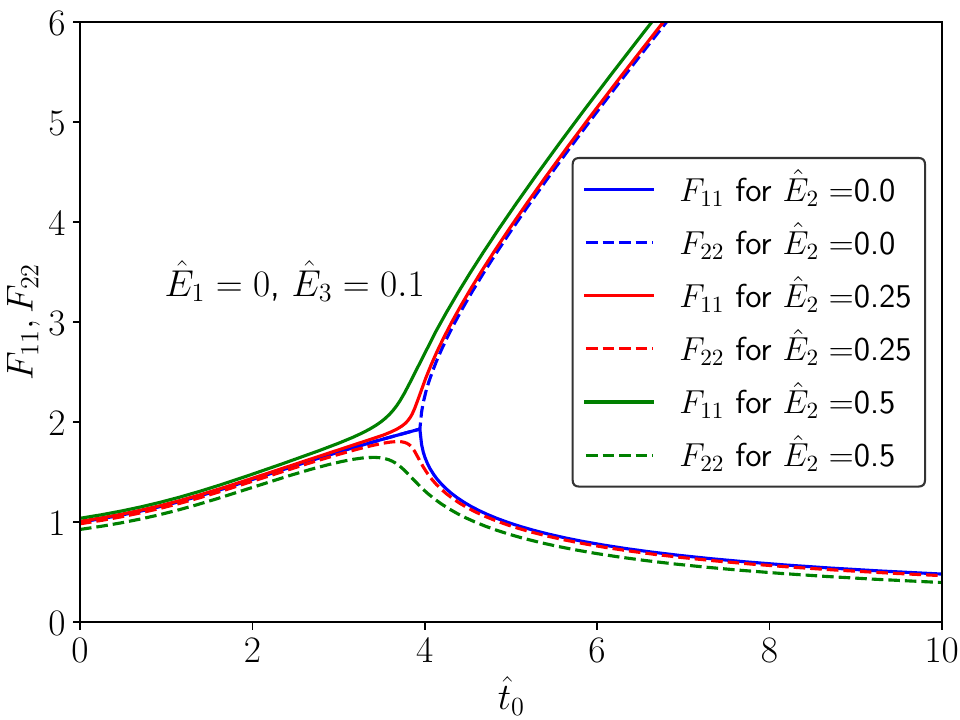} \label{fig:bif2}}
    \\	
    \subfloat[]{\includegraphics[width=0.49\textwidth]{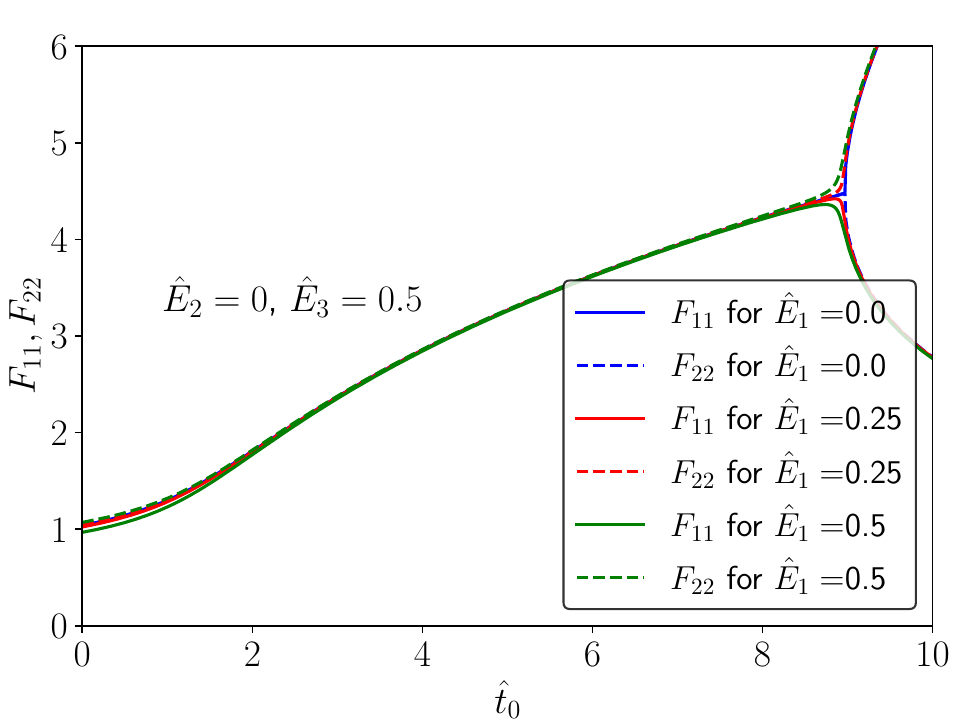} \label{fig:bif3}}
    \caption{
        Bifurcation diagram of the DE disk. 
        (a) and (b) show that the effect of the in-plane field on the bifurcation behavior is minimal in terms of modifying the bifurcation load, but they are essential in breaking the symmetry by nudging the circular disk to adopt an elliptical cross-section whose smaller principal axis is aligned with the applied in-plane field.
        (c) shows the much more significant impact of the out-of-plane field in changing the bifurcation structure.
    }
\end{figure}

The stability of the dielectric elastomer is studied by using the energy of the system to derive conditions under which a bifurcated solution will occur. The implicit function theorem provides the condition for bifurcation to occur by determining the solutions for a vanishing Hessian matrix. For our numerical results, we will minimize the energy density in \eqref{eq:comp_engy} for various loading cases to study the deformation while keeping the axes in the current configuration fixed. 

Figure \ref{fig:bif1} shows the effect of the electric field on the deformation. 
Initially, at $\hat{t}_0 = 0$, we observe that $(F_{11}, F_{22}) = (1,1)$ when $\hE_1 = 0$, $(F_{11}, F_{22}) = (0.98,1.01)$ when $\hE_1 = 0.25$ and $(F_{11}, F_{22}) = (0.93,1.04)$ when $\hE_1 = 0.5$. 
We highlight that $\hE_1$ introduces a asymmetry in stretches between the $1-$ and $2-$ directions. 
As $\hat{t}_0$ increases, both $F_{11}$ and $F_{22}$ increase while maintaining the asymmetry at a small value below a critical load. 
The solution, then, bifurcates when the load exceeds the critical value and the asymmetry in the stretches increases dramatically. 
We note that for $\hE_1= \hE_2 = 0$, there is an unstable solution where $F_{11} = F_{22}$ as demonstrated in \cite{kearsley1986asymmetric,chen2021interplay}. 
However, when we introduce in-plane components of electric field, this acts to perturb the symmetrical state of the dielectric, leading to only bifurcated solutions. 

In Figure \ref{fig:bif2}, the orientation of the in-plane electric field is changed to be aligned along the $2-$direction. 
We observe the same behavior as Figure \ref{fig:bif1}, but with magnitude reversed between $F_{11}$ and $F_{22}$, showing the clear role of the in-plane electric field in biasing the bifurcated configuration.

Figure \ref{fig:bif3} shows the effect of changing the magnitude of the out-of-plane field on the bifurcation behavior. 
We observe that for $\hE_3=0.5$, the TK instability is delayed to a higher load value and the effect of $\hE_1$ on the deformation is reduced.

\begin{figure}[htb!]
    \centering
    \includegraphics[width=0.49\textwidth]{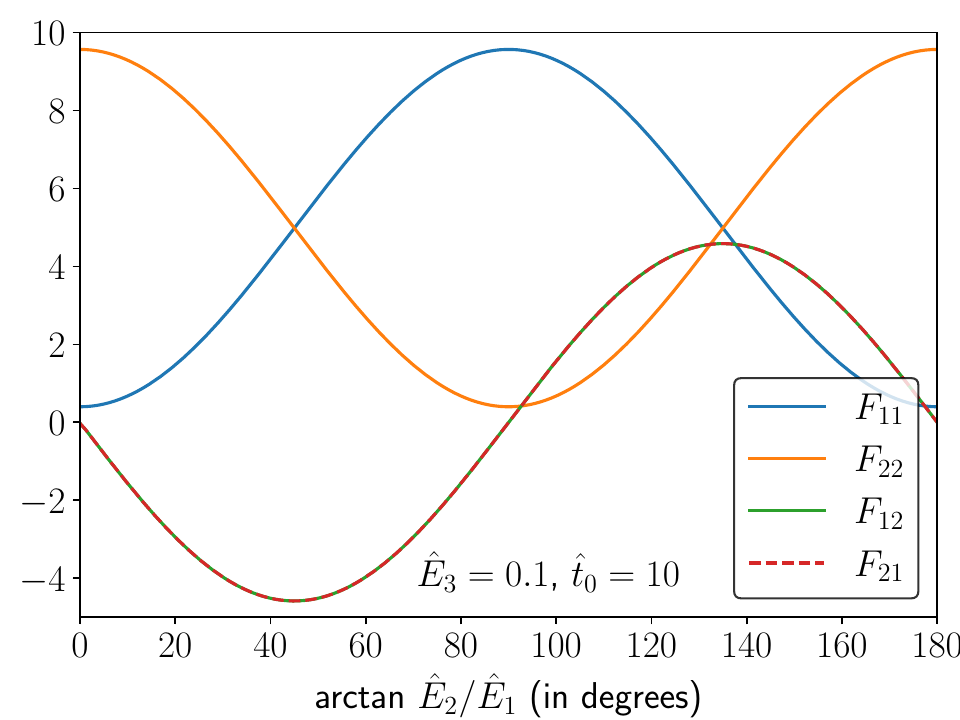} 
    \caption{
        The variation of the components of $\bfF$ with the orientation of the in-plane electric field in the bifurcated state.
        We notice that the circular disk takes on an elliptical cross-section with the minor principal axis aligned with the electric field.
        We highlight that $F_{12} \neq F_{21}$ but their numerical values are very close to each other, which indicates that the disk is close to, but not perfectly, elliptical.
    }
\label{fig:bif4}
\end{figure}

In Figure \ref{fig:bif4}, we plot the change in the components of $\bfF$ with respect to the orientation of the electric field, denoted as $\alpha := \arctan{\hE_2/\hE_1}$, while keeping the magnitude of the electric field fixed at $0.1$. 
The mechanical load is fixed at $\hat{t}_0 = 6$ and the through-thickness component the electric field is fixed at $\hE_3 = 0.1$; note that these values put the DE disk in the bifurcated state.
As $\alpha$ changes, we can solve for $\bfF$ from \eqref{str1}; we find that the deformation changes such that the minor principal axis rotates to follow $\alpha$, and the specimen contracts in the direction of the electric field and extends in the orthogonal direction.
This is consistent with Figures \ref{fig:bif1} and \ref{fig:bif2} that correspond to $\alpha=0^\circ$ and $\alpha=90^\circ$ respectively.

\section{Extension to Realistic Complex Geometries} \label{fem}

While the theoretical analysis above provides essential insights in an idealized setting, we complement it by looking at a more realistic geometry that is solved using FEM.
In particular, the assumption that the applied boundary potential has the form $\phi = -\bfE^{ext}\cdot\bfx$ is essentially impossible to realize in practice.
Therefore, we consider a specimen that has several distinct electrodes on the boundary, each of which is held at a given voltage (Fig. \ref{fig:figbc}).
By changing the voltage in each electrode as in Figure \ref{fig:voltbc}, we are able to effectively rotate the overall electric field vector to achieve the proposed actuation mechanism.
Further, we highlight that the electrodes are on the top and bottom surfaces rather than along the thin lateral surface, in the interest of providing a geometry that is potentially feasible for experimental realization.

\subsection{Patterned Surface Electrode Geometry and Voltages}

\begin{figure}[htb!]
    \centering
    \subfloat[]{\includegraphics[width=0.48\textwidth]{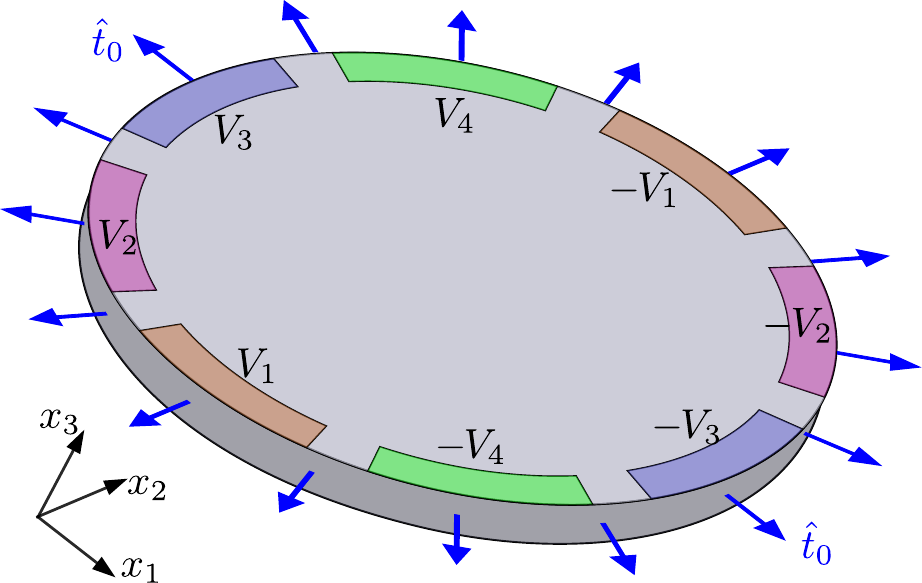} \label{fig:figbc}}
    \hfill
    \subfloat[]{\includegraphics[width=0.48\textwidth]{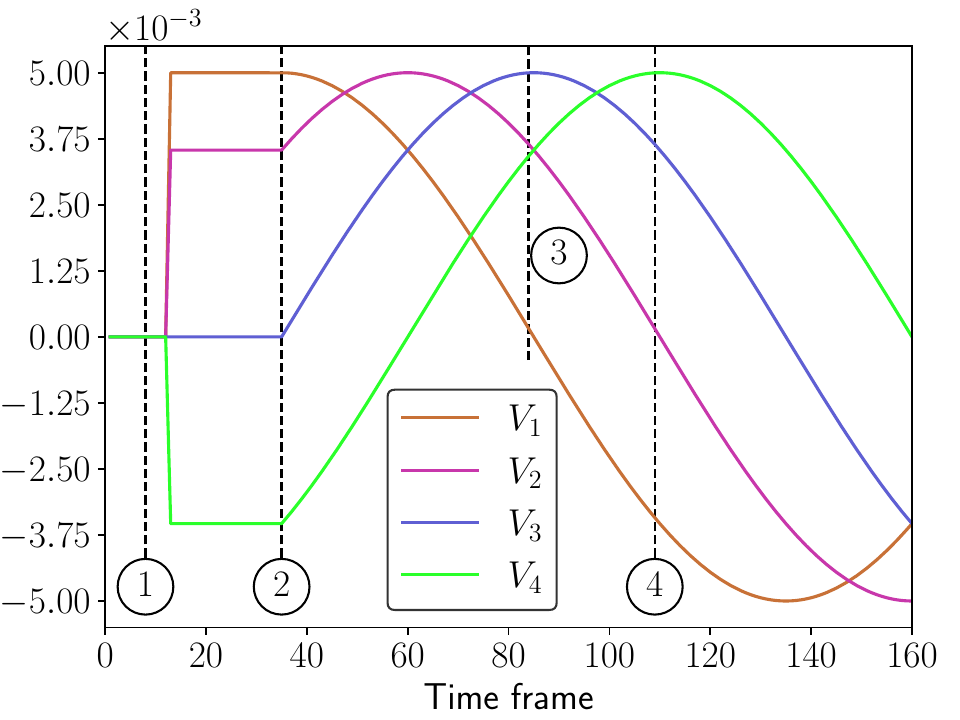} \label{fig:voltbc}}
    \caption{
        (a) The geometry of the DE specimen showing the applied traction and the patterned electrodes on the top and bottom surfaces. 
        (b) The voltages applied to the various electrodes as a function of time. However, we emphasize that the variation in time is strictly for conceptual convenience; we do not consider any dynamic phenomena, and solve for equilibrium at each instant as the voltage boundary conditions evolve.
    }
\end{figure}

As shown in Figure \ref{fig:figbc}, we consider a circular specimen with a uniform thickness and a uniform applied traction on the boundary.
We restrain the body only to prevent rigid modes.
The voltage BC are applied non-uniformly on the top and bottom surfaces through an arrangement of eight electrodes, and the voltages that are assigned to the top electrodes are denoted in Figure \ref{fig:figbc}.
The same voltages are assigned on the bottom boundary. 
The remainder of the boundary has no electrodes. 
While we choose 8 electrodes for convenience, it straightforward to have a larger number.

The sequence of loading is as follows:
\begin{enumerate}
    \item The mechanical load $\hat{t}_0 = 3.86$ is applied while the voltage is held at $0$ at all electrodes, over the time interval $t = 0$ to $t=12$ in Figure \ref{fig:voltbc}.
    \item A constant nonzero voltage is applied to the electrodes as shown in Figures \ref{fig:figbc} and \ref{fig:voltbc}, over the time interval $t=12$ to $t=34$. The mechanical load is increased to $\hat{t}_0=4.5$ within this time interval.
    \item The mechanical load is held fixed at $\hat{t}_0=4.5$ throughout the rest of the calculation, while the voltages at the electrodes are varied sinusoidally with each electrode having a different phase.
\end{enumerate}

The equilibrium BVP defined by \eqref{Maxwell}, \eqref{equil1}, \eqref{equil2}, and \eqref{equil3}, with the BC described above, is solved using the open source FEM package FEniCS \cite{alnaes2015fenics,logg2010dolfin}. 
We used a triangular mesh with quadratic shape functions for the displacement and electric field and linear shape functions for the Lagrange multiplier. 
We performed a static three-dimensional analysis with the voltage/load applied incrementally.

Because the overall system is nonlinear with a constraint, we solve as follows.
At each increment (either traction or voltage), the electrostatic problem \eqref{Maxwell} is solved first. 
The electric field thus obtained is used to solve the mechanical problem \eqref{equil1}, \eqref{equil2} and \eqref{equil3}.
The mechanical quantities are used to solve the electrostatic problem, and the entire process is repeated until convergence.

\subsection{Results and Discussion}

\begin{figure}[htb!]
    \centering
	\subfloat[\label{fig:snapshot1}]{%
      \begin{overpic}[width=0.49\textwidth]{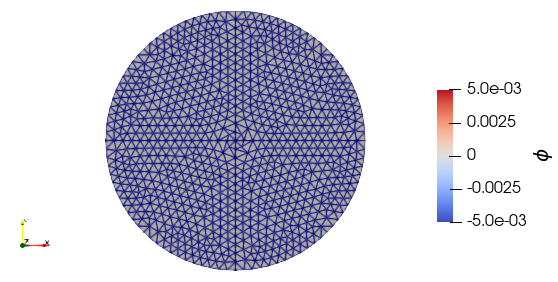}
        \put(12,60){\color{black}\small \Circled{1}}
        \put(12,50){\color{black}\small $\hat{t}_0 = 0.54$}
      \end{overpic}
    }
	\hfill
	\subfloat[\label{fig:snapshot2}]{%
      \begin{overpic}[width=0.49\textwidth]{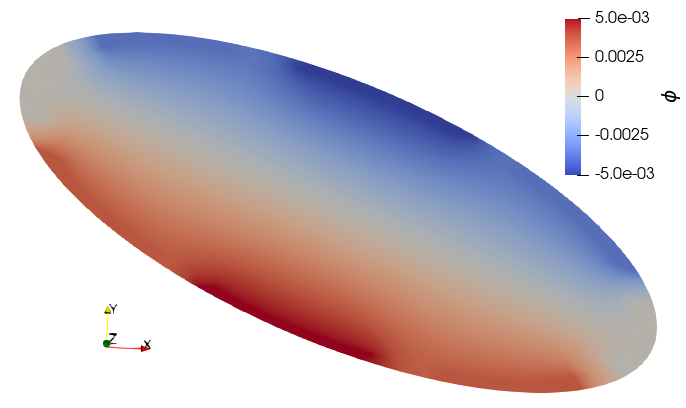}
        \put(60,60){\color{black}\small \Circled{2}}   
        \put(60,50){\color{black}\small $\hat{t}_0 = 4.5$}
      \end{overpic}
    }
	\\
	\subfloat[\label{fig:snapshot3}]{%
      \begin{overpic}[width=0.45\textwidth]{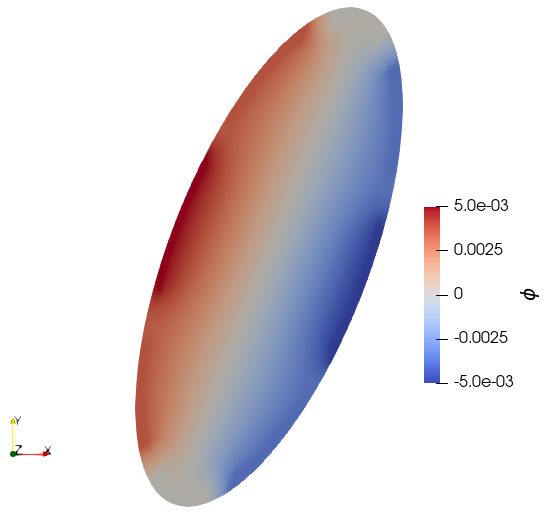}
        \put(25,80){\color{black}\small \Circled{3}}
        \put(25,70){\color{black}\small $\hat{t}_0 = 4.5$}
      \end{overpic}}
	\hfill
	\subfloat[\label{fig:snapshot4}]{%
      \begin{overpic}[width=0.49\textwidth]{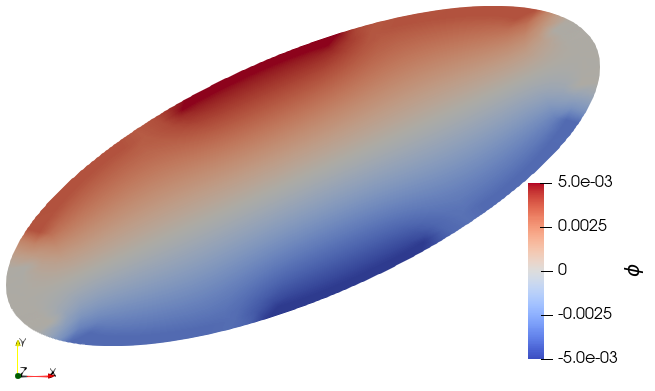}
        \put(20,70){\color{black}\small \Circled{4}}
        \put(20,60){\color{black}\small $\hat{t}_0 = 4.5$}
      \end{overpic}
    }
    \caption{
        Snapshots of the deformed specimen as the electrode voltage evolves; the labels \Circled{1}-\Circled{4} indicate the times that are shown in Figure \ref{fig:voltbc}. The color indicates the electric potential. In (a), we have $\hat{t}_0 = 0.54$ which is below the bifurcation load, and in (b), (c), (d), we have $\hat{t}_0 = 4.5$ which is above the bifurcation load.
    }
\end{figure}

\begin{figure}[htb!]
    \centering
	\subfloat[\label{fig:elsnapshot1}]{%
      \begin{overpic}[width=0.49\textwidth]{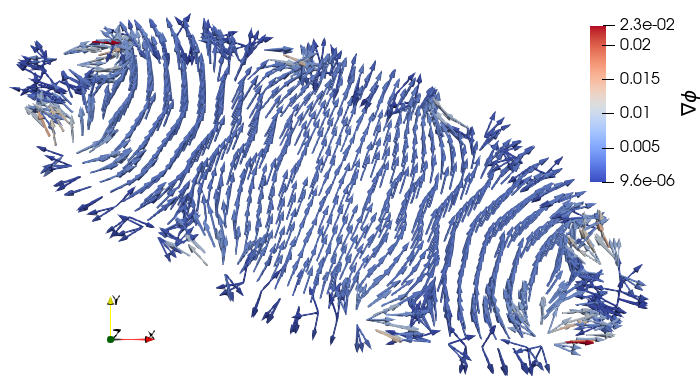}
        \put(12,75){\color{black}\small \Circled{2}}
        \put(12,65){\color{black}\small $\hat{t}_0 = 4.5$}
      \end{overpic}
    }
	\hfill
	\subfloat[\label{fig:elsnapshot2}]{%
      \begin{overpic}[width=0.4\textwidth]{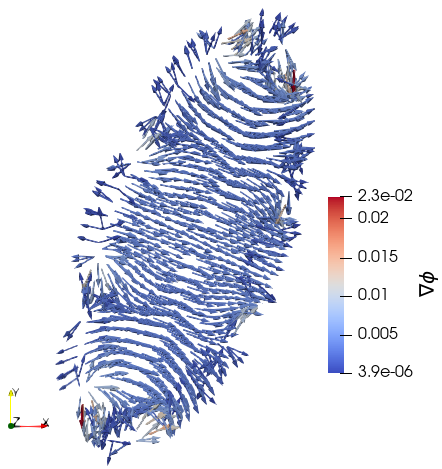}
        \put(12,85){\color{black}\small \Circled{3}}   
        \put(12,75){\color{black}\small $\hat{t}_0 = 4.5$}
      \end{overpic}
    }
    \caption{Top view of electric field in the specimen at \Circled{2} and \Circled{3}.}
\end{figure}

Figure \ref{fig:snapshot1} shows the deformed shape of the specimen at a mechanical load $\hat{t}_0 = 0.54$ at time \Circled{1} (Fig. \ref{fig:voltbc}) with zero voltage applied at the electrodes. 
This load is below the bifurcation load of $\hat{t}_0 = 3.89$ for the TK instability in the absence of electric field  (Fig. \ref{fig:bif1}). 
Thus, the specimen expands symmetrically into a circle with larger radius in the plane while reducing its thickness. 

As we increase the mechanical load to $\hat{t}_0 = 4.5$, the voltages $V_1, V_2, V_3$ and $V_4$ are prescribed on the electrodes at time $t=12$ and held constant up to $t=35$ (Fig. \ref{fig:voltbc}).
The electric field is no longer uniform, the maximum magnitude is $2.3 \times 10^{-2}$, while the field in most of the material is around $3 \times 10^{-3}$, which is far less than those in the analysis in Section \ref{Analysis_Results}. Thus, the TK instability will occur very close to $\hat{t}_0 = 3.89$ which is the value for the purely mechanical case.

In Figure \ref{fig:snapshot2}, at the time \Circled{2}, we have $\hat{t}_0 = 4.5$ which is above the bifurcation load and the specimen is clearly elliptical.
Further, the electrodes set up an electric field within the specimen that is fairly uniform (Fig. \ref{fig:elsnapshot1}), and the major axis of the ellipse is clearly aligned with the overall  electric field.

Subsequently, as the electrode voltage evolves while holding fixed the mechanical load at $\hat{t}_0 = 4.5$, the maximum  and minimum voltage gradually change position; Figure \ref{fig:snapshot3} shows this at $t=84$. 
The direction of the in-plane electric field also changes accordingly (Fig. \ref{fig:elsnapshot2}), which in turns causes the major axis of the specimen to rotate. 

A similar result is shown in Figure \ref{fig:snapshot4} at $t=109$.

\section{Multilayered Dielectric Elastomer Actuators}
\label{sec:multi-layer}

The design of Dielectric Elastomers (DE) for electromechanical actuation has included multilayered DEs stacked to enhance stability \cite{duduta2016multilayer} and enabling actuation without applying a prestretch. 
We analyze a simple multilayer configuration to demonstrate that the electromechanical TK effect provides a useful actuation mechanism in this setting as well.

\begin{figure}[htb!]
    \centering
	\subfloat[\label{fig:multilayer1}]{%
     \includegraphics[width=0.25\textwidth]{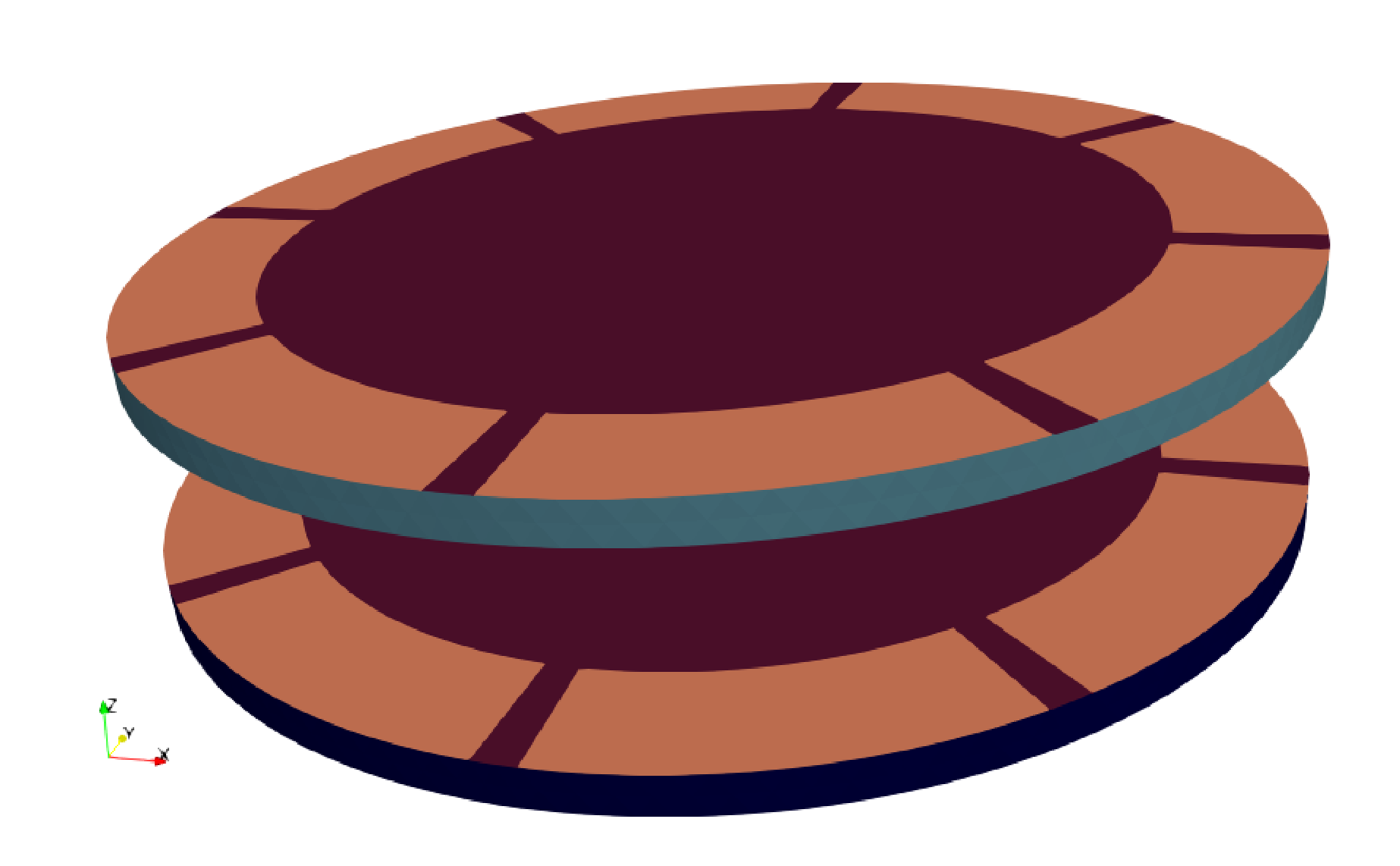}}
	\subfloat[\label{fig:multilayer2}]{%
      \includegraphics[width=0.25\textwidth]{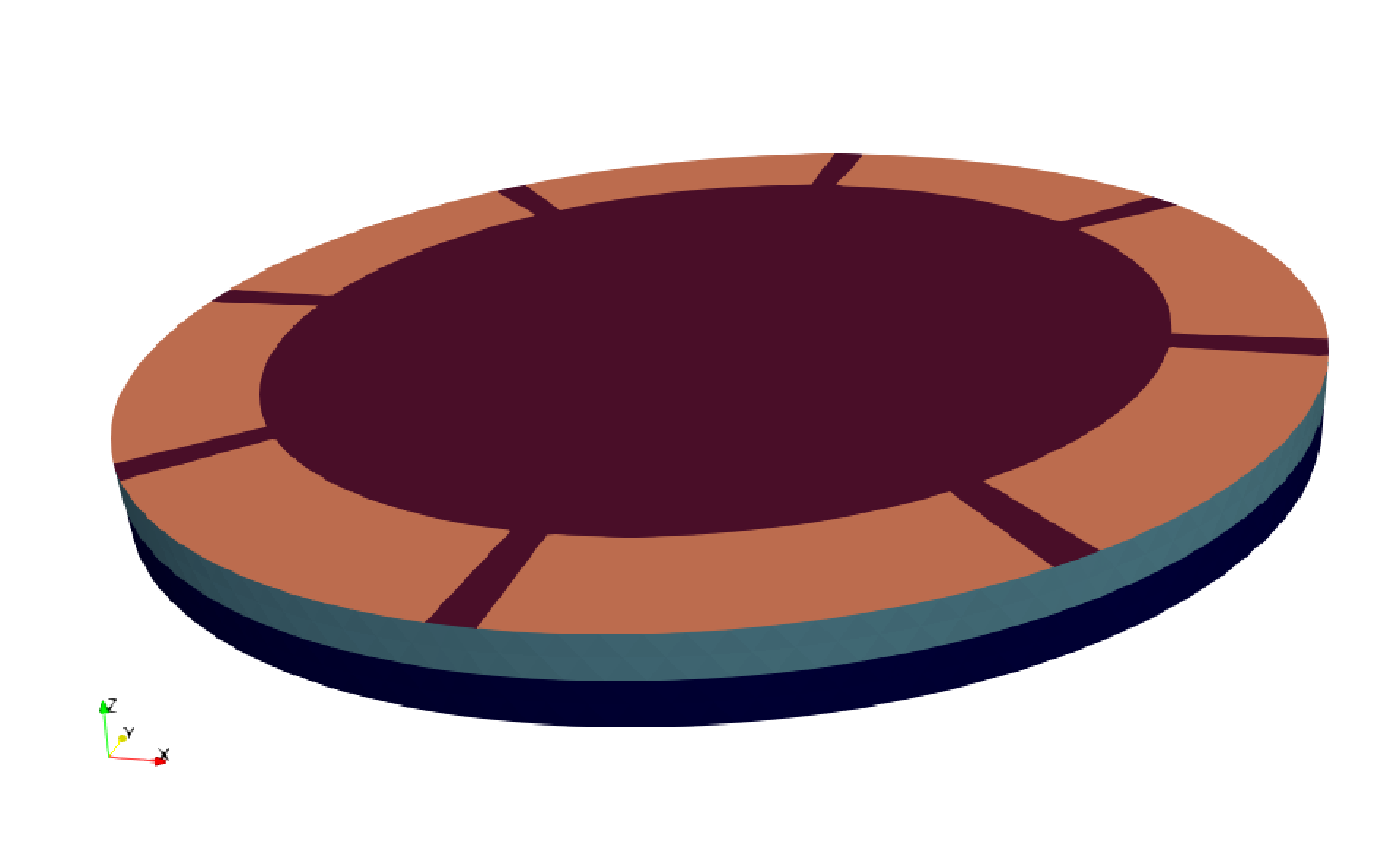}
    }
    \caption{Two-layer Dielectric Elastomer: (a) a schematic showing the electrodes in the interface between the layers, (b) the computed stack assembled such that electrodes in contact have identical voltages.}
\end{figure}

In our analysis, we stack two DEs with the electrodes arranged in the pattern shown in Figures \ref{fig:multilayer1} and \ref{fig:multilayer2}.
The arrangement of electrodes in each layer is similar to the single-layer configuration (Fig. \ref{fig:figbc}), and we ensure that electrodes with the same voltage are connected at the interface. 
The FEM calculations use similar boundary conditions as those of the single later, while enforcing continuity of the displacement at the interface. 
A uniform traction in applied to both layers around the circumference.

\begin{figure}[htb!]
    \centering
	\subfloat[\label{fig:Multilayer_voltage1}]{%
      \begin{overpic}[width=0.45\textwidth]{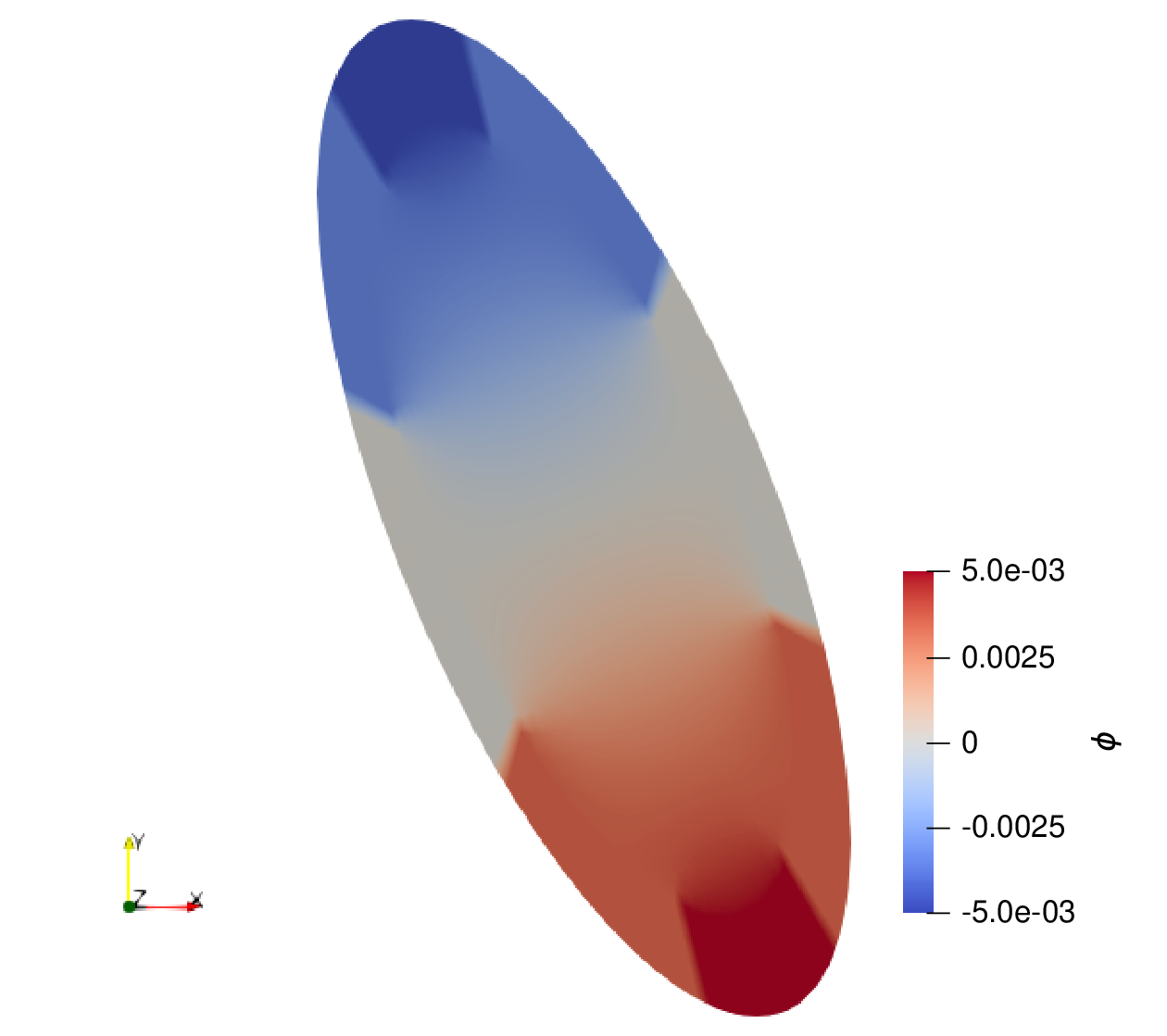}
        \put(12,25){\color{black}\small $\hat{t}_0 = 4.5$}
      \end{overpic}
    }
	\hfill
	\subfloat[\label{fig:Multilayer_voltage2}]{%
      \begin{overpic}[width=0.45\textwidth]{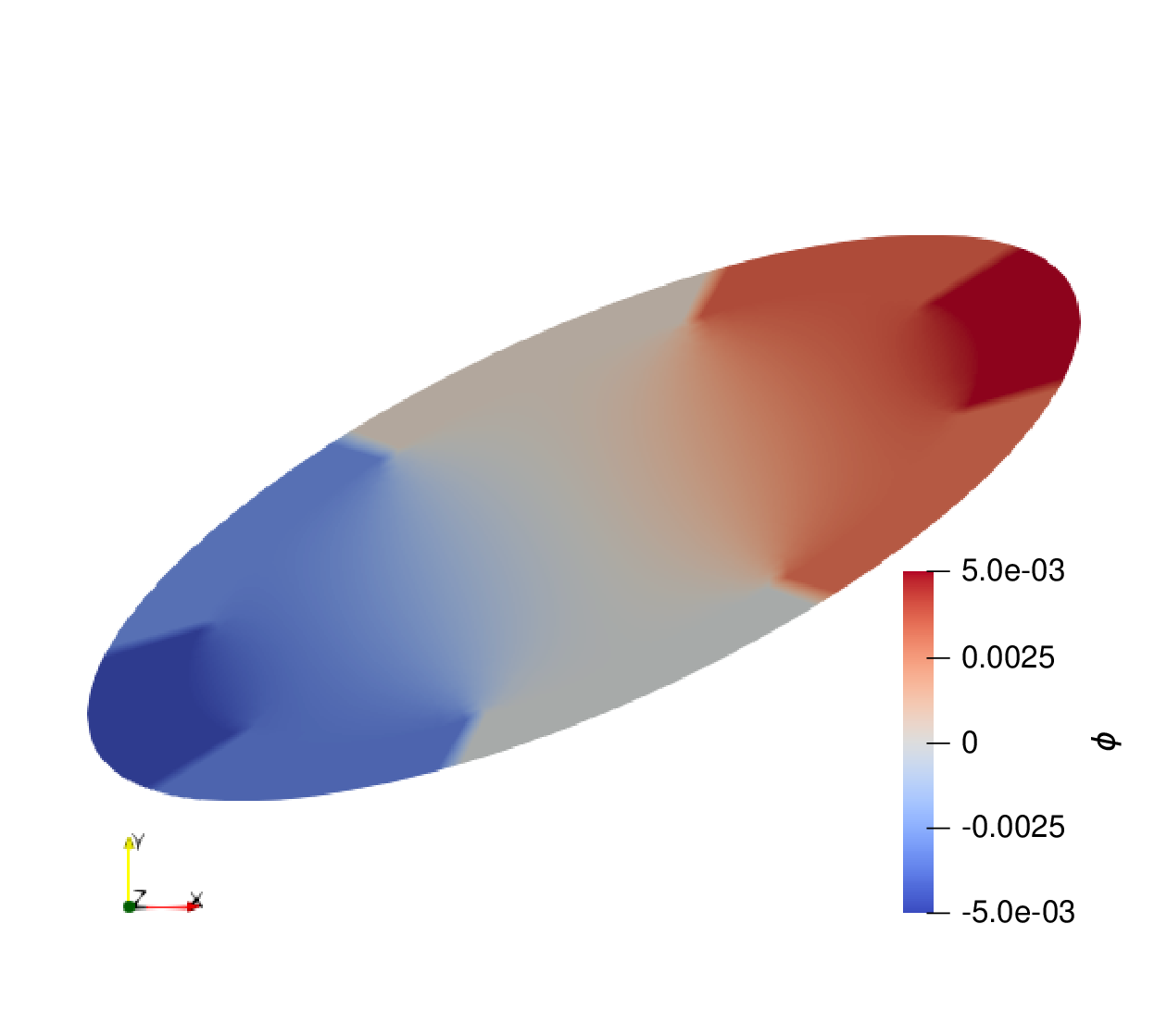}
        \put(20,60){\color{black}\small $\hat{t}_0 = 4.5$}
      \end{overpic}
    }
	\\
	\subfloat[\label{fig:Multilayer_efield1}]{%
      \begin{overpic}[width=0.45\textwidth]{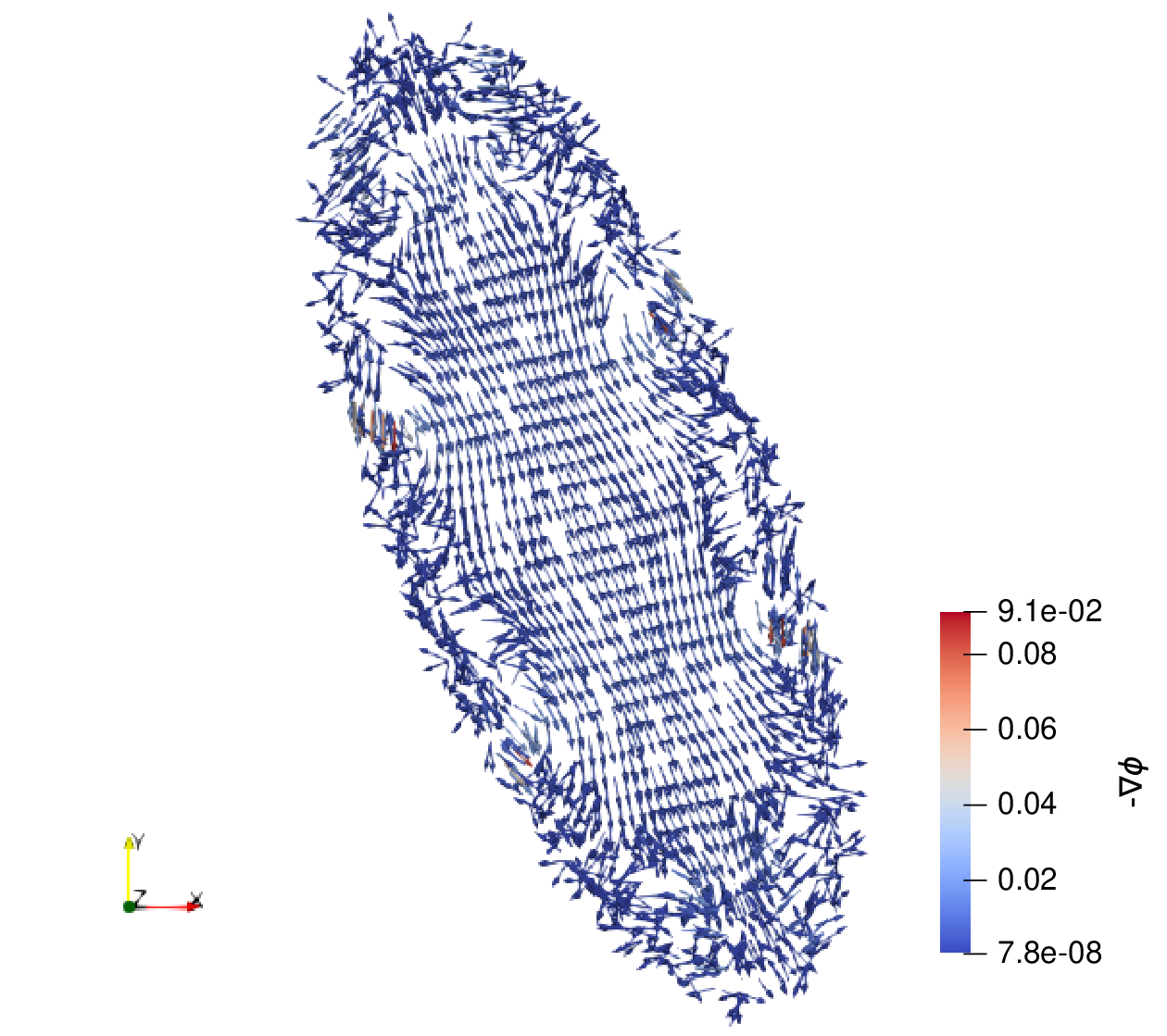}
        \put(12,25){\color{black}\small $\hat{t}_0 = 4.5$}
      \end{overpic}}
	\hfill
	\subfloat[\label{fig:Multilayer_efield2}]{%
      \begin{overpic}[width=0.49\textwidth]{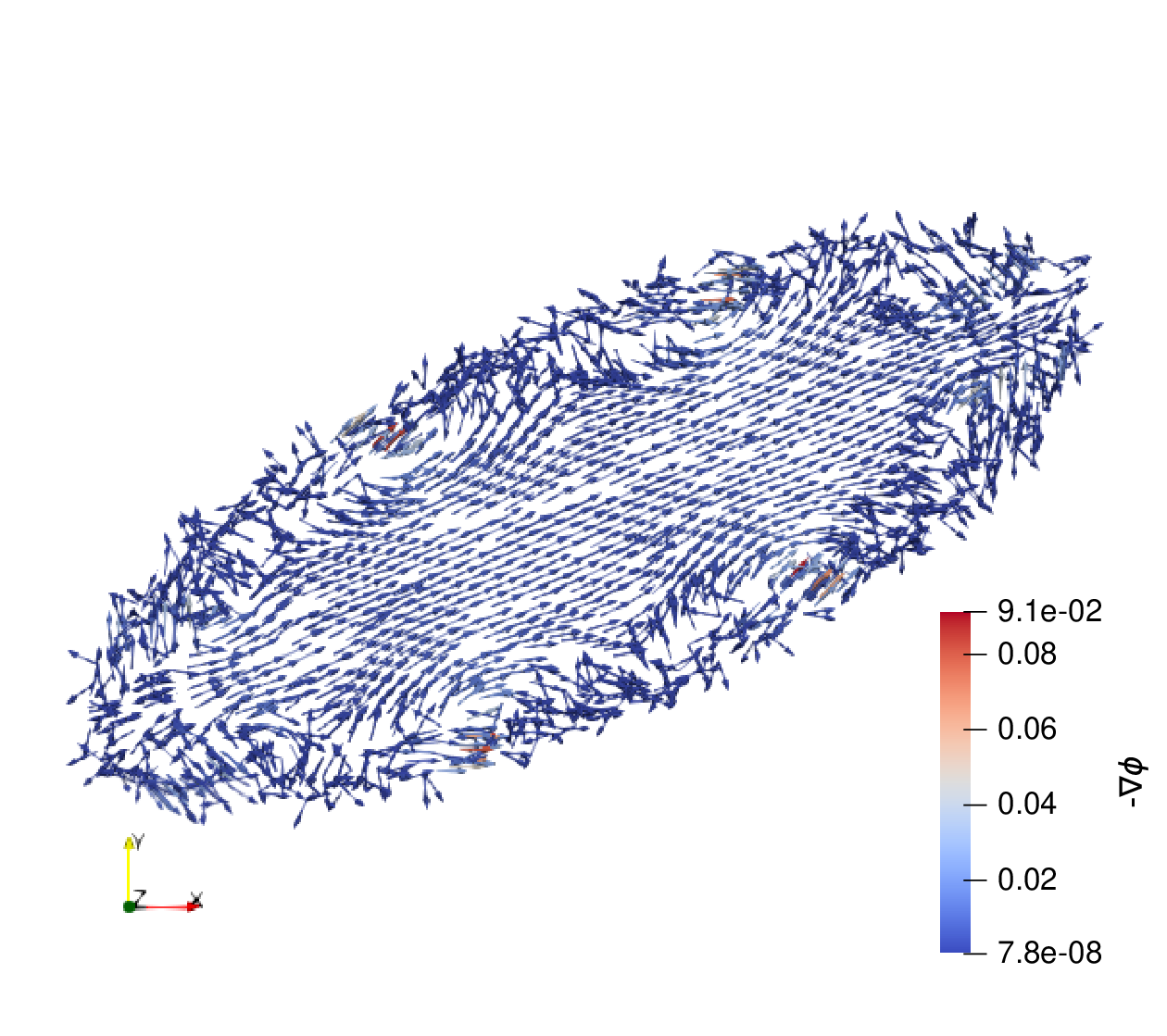}
        \put(20,60){\color{black}\small $\hat{t}_0 = 4.5$}
      \end{overpic}
    }
    \caption{
        Snapshots of the deformed specimen as the electrode voltage evolves (Fig. \ref{fig:figbc}) in the two-layer DE at a load $\hat{t}_0 = 4.5$ in a post-bifurcation state. In (a) and (b), the color indicates the electric potential while in (c) and (d), the color indicates the electric field magnitude.
    }
\end{figure}

The FEM results show that the traction at bifurcation is $\hat{t}_0 = 3.89$, which is the same as in the single layer.
We apply the same voltages to the electrodes as in the case of the single layer DE (Fig. \ref{fig:voltbc}).
The post-bifurcation deformation in Figures \ref{fig:Multilayer_voltage1} and \ref{fig:Multilayer_voltage2} has its principal direction oriented in the same direction as the electric field in the middle of the DEs shown in Figures \ref{fig:Multilayer_efield1} and \ref{fig:Multilayer_efield2}. 

We highlight an interesting difference between single-layer and multilayer DEs: the single-layer DEs deformed such that the short axis was aligned with the average electric field, whereas in multilayer DEs the long axis is aligned with the average electric field.
The reason for this difference is as follows.
In single-layer DEs, the alignment is dominated by Coulombic attraction across the specimen: the electrodes with the largest voltages attract and squeeze the specimen, causing the short axis to align with the field.
On the other hand, in multilayer DEs, the electric fields at the edges of the specimen between the patterned electrodes are dominate over the average field in the interior of the specimen in determining the response, and these local fields drive the long axis to align with the electric field.
The interactions between adjacent electrodes is amplified due to the multilayer stacking configuration.
This is tested by changing the spacing and number of electrodes.

We highlight that regardless of the orientation of the deformation --- whether the long axis or the short axis aligns with the electric field --- the key idea of exploiting the electromechanical TK instability is unchanged.
That is, we have a flat energy landscape and rotating the electric field, by changing the voltages on the electrodes, causes the system to traverse  the flat landscape and exhibit large shape changes with negligible work.

\section{Concluding Remarks} 
\label{concl}

We have presented a strategy to achieve large and soft deformations in dielectric elastomers at low voltages by exploiting the TK instability.
In particular, the TK instability is a symmetry-breaking bifurcation that provides an energy landscape that has a significant flat region.
The flat region allows for symmetry-related soft modes and it costs no energy to move over this region of the energy landscape.
Consequently, by applying a small electric field that appropriately biases the system, we show the possibility of nudging the system to a desired configuration.
We demonstrated this in an idealized system theoretically as well as numerically in realistic single-layer and multilayer DE configurations.

Our strategy of working near a pitchfork bifurcation is analogous to exploiting a second-order phase transition \ref{fig:idea}.
If one could conceive of an approach that exploits a first-order phase transition, it could be possible to achieve much larger field-induced deformations, although possibly introducing hysteresis.
Woven structures are a model setting to potentially achieve a first-order transition using equibiaxial boundary traction \cite{liu1995stability}, similar to the load case studied in this paper.



\paragraph*{Software and Data Availability.}

The code developed for this work and the associated data are available at \\ \url{https://github.com/danielkatusele/Dielectric-Shape}

\paragraph*{Acknowledgments.}

We thank NSF (DMS 2108784, DMREF 1921857), BSF (2018183), and AFOSR (MURI FA9550-18-1-0095) for financial support; the TCS Presidential Fellowship to Daniel Katusele for additional support; NSF for XSEDE computing resources provided by Pittsburgh Supercomputing Center; Timothy Breitzman and Matthew Grasinger for useful discussions; and the anonymous reviewers to pointing us to the literature on multilayer dielectric elastomers.

\end{document}